\documentclass[%
 reprint,
superscriptaddress,
 amsmath,amssymb,
 aps,
]{revtex4-2}
\usepackage{adjustbox}
\usepackage{graphicx}
\usepackage{dcolumn}
\usepackage{bm}
\usepackage{adjustbox}
\usepackage{gensymb}
\usepackage{float}
\usepackage[OT4]{fontenc}
\usepackage[mathlines]{lineno}

\begin{document}

\preprint{APS/123-QED}

\title{Highly ${ }^{28} \mathrm{Si}$ Enriched Silicon  by Localised Focused Ion Beam Implantation}

\author{Ravi Acharya}
\email{ravi.acharya@manchester.ac.uk}
\affiliation{Department of Electrical and Electronic Engineering, Photon Science Institute, University of Manchester, Manchester, M13 9PL, UK}
\affiliation{School of Physics, University of Melbourne, Parkville, Melbourne, VIC 3010, Australia }
\author{Maddison Coke}
\affiliation{Department of Electrical and Electronic Engineering, Photon Science Institute, University of Manchester, Manchester, M13 9PL, UK}
\author{Mason Adshead}
\affiliation{Department of Electrical and Electronic Engineering, Photon Science Institute, University of Manchester, Manchester, M13 9PL, UK}
\author{Kexue Li}
\affiliation{Department of Materials, Photon Science Institute, University of Manchester, Manchester, M13 9PL, UK}
\author{Barat Achinuq}
\affiliation{Department of Materials, University of Manchester, Manchester, M13 9PL, UK}
\author{Rongsheng Cai}
\affiliation{Department of Materials, University of Manchester, Manchester, M13 9PL, UK}
\author{A. Baset Gholizadeh}
\affiliation{Department of Electrical and Electronic Engineering, Photon Science Institute, University of Manchester, Manchester, M13 9PL, UK}
\author{Janet Jacobs}
\affiliation{Department of Electrical and Electronic Engineering, Photon Science Institute, University of Manchester, Manchester, M13 9PL, UK}
\author{Jessica L. Boland}
\affiliation{Department of Electrical and Electronic Engineering, Photon Science Institute, University of Manchester, Manchester, M13 9PL, UK}
\author{\\Sarah J. Haigh}
\affiliation{Department of Materials, University of Manchester, Manchester, M13 9PL, UK}
\author{Katie L. Moore}
\affiliation{Department of Materials, Photon Science Institute, University of Manchester, Manchester, M13 9PL, UK}
\author{David N. Jamieson}
\affiliation{School of Physics, University of Melbourne, Parkville, Melbourne, VIC 3010, Australia }
\author{Richard J. Curry}
\email{richard.curry@manchester.ac.uk}
\affiliation{Department of Electrical and Electronic Engineering, Photon Science Institute, University of Manchester, Manchester, M13 9PL, UK}

\date{\today}

\begin{abstract}
Solid-state spin qubits within silicon crystals at mK temperatures show great promise in the realisation of a fully scalable quantum computation platform. Qubit coherence times are limited in natural silicon owing to coupling to the isotope ${ }^{29} \mathrm{Si}$ which has a non-zero nuclear spin. This work presents a method for the depletion of ${ }^{29} \mathrm{Si}$ in localised volumes of natural silicon wafers by irradiation using a 45 keV ${ }^{28} \mathrm{Si}$ focused ion beam with fluences above $1 \times 10^{19} \, \mathrm{ions} \, \mathrm{cm}^{-2}$.  Nanoscale secondary ion mass spectrometry analysis of the irradiated volumes shows unprecedented quality enriched silicon that reaches a minimal residual ${ }^{29} \mathrm{Si}$ value of 2.3 $\pm$ 0.7  ppm and with residual C and O  comparable to the background concentration in the unimplanted wafer. Transmission electron microscopy lattice images confirm the solid phase epitaxial re-crystallization of the as-implanted amorphous enriched volume extending over 200 nm in depth upon annealing. The ease of fabrication, requiring only commercially available natural silicon wafers and ion sources, opens the possibility for co-integration of qubits in localised highly enriched volumes with control circuitry in the surrounding natural silicon for large-scale devices.
\end{abstract}
\maketitle


\section{Introduction}
Development of device architectures for a large-scale quantum computer are underway worldwide \cite{General_adv_mat}. A crucial issue for all these approaches is the need to develop  robust qubits that allow high-fidelity operation. Coherent quantum systems are highly-sensitive to their local environment which leads to qubit decoherence.  This necessitates the use of quantum error correction codes \cite{takeda2022quantum} which impose significant engineering overheads such that millions of qubits are required to realise a fault-tolerant quantum computer \cite{saraiva2022materials}.\\

Spin qubits in silicon present a promising platform to realise scalable quantum computation \cite{kane1998Silicon} due to their long coherence times in the solid-state and high gate fidelities \cite{Semiconductorqubitsinpractice, muhonen2014storing, dehollain2016optimization}, as well as compatibility with the materials and processes employed for industrial nano-fabrication techniques \cite{CMOScompatbilitySi}. However, in natural silicon (${ }^{\mathrm{Nat}}\mathrm{Si}$), the nuclear spin \textit{I} = 1/2 \cite{29Spin} ${ }^{29}\mathrm{Si}$ isotope with abundance 4.68\% \cite{Abundances} acts to fundamentally limit electron spin qubit coherence times in donor and dot architectures. This is due to the dipole-induced ${ }^{\mathrm{29}}\mathrm{Si}$ nuclear spin flip-flops within the nuclear spin bath surrounding the electron spin \cite{Sousa2003, Witzel2006, Witzel2010}. In isotopically enriched ${ }^{\mathrm{28}}\mathrm{Si}$ (\textit{I} = 0) electron spin qubit coherence times can be significantly increased.  For example, a substrate with 800 ppm residual ${ }^{29}\mathrm{Si}$ has led to a factor of 5000 improvement in the electron pure dephasing time $(\textit{T}_{2e}^{*})$ as compared to ${ }^{\mathrm{Nat}}\mathrm{Si}$ \cite{muhonen2014storing}. A matrix composed of a single isotope is also preferable because lattice strain caused by isotopic mass variations in the crystal lattice can perturb the hyperfine interaction between the donor nucleus and electron \cite{itoh_watanabe_2014}. 
It is also important to minimise the background concentration of impurities and lattice defects that may couple to the qubit. \\

Several techniques have been applied to produce enriched ${ }^{28} \mathrm{Si}$ (\textbf{Table \ref{tab:tablesummary}}). 
The highest quality material regarding both enrichment and contamination was produced for the ``Avogadro project'' \cite{Avogadro}, where centrifugation was employed to isotopically separate natural silicon tetraflouride gas ${ }^{\mathrm{Nat}}\mathrm{SiF_{4}}$ in order to produce ${ }^{\mathrm{28}}\mathrm{SiF_{4}}$. The combined background  ${ }^{30}\mathrm{Si}$ and  ${ }^{29}\mathrm{Si}$ isotopic level of $\sim$ 50 ppm was later improved upon by the  ``kg-2 project'' \cite{kg-2} which reported residual ${ }^{29}\mathrm{Si}$ and  ${ }^{30}\mathrm{Si}$ levels of 6.6 ppm and 0.38 ppm, respectively. Ion beams that employ mass-filtering using magnetic and electric fields can be used to create isotopically enriched material. An example of this procedure is the creation of epitaxial enriched thin films via hyperthermal ion beam deposition employing a Penning ion source \cite{ dwyer2014enriching, Hypothermal1, tang2020targeted} from a natural silane source gas. However, the material is reported to contain a significant level of C and O greater than the backgrounds found in device grade silicon wafers.\\
\begin{table*}
\caption{\label{tab:tablesummary} A summary of the reported ${ }^{29}\mathrm{Si}$ and ${ }^{30}\mathrm{Si}$ depletion level, and background contaminant level, reached by a variety of different enrichment techniques. The results outlined in this work employing the use of a focused ion beam are also shown. Each value is quoted to the same precision that was reported.
}
\begin{adjustbox}{width=1\linewidth}
\begin{ruledtabular}
\begin{tabular}{cccccc}
 Reference & Method & ${}^{29}\mathrm{Si}$ (ppm) & ${}^{30}\mathrm{Si}$ (ppm) & C  ($\mathrm{cm}^{-3}$)  & O ($\mathrm{cm}^{-3}$)\\
    \hline
     \noalign{\vskip 0.03in} 
``Avogadro project''  \cite{Avogadro} & Centrifuge  &   $<$ 50.65 & $<$ 50.65 & $(3.4 \, \pm  \,4.0) \times 10^{14}$ & $(2.1 \, \pm 1) \times 10^{14}$  \\
``kg-2 project''  \cite{kg-2} & Centrifuge & 6.583 $\pm$ 0.031 & 0.378 $\pm$ 0.010 & $ <4 \times 10^{15}$ & $<5\times 10^{14}$\\ 
Mazzocchi et al. \cite{CVDMAZZOCCHI20191} & Centrifuge  & 52.4 $\pm$ 9 & 13.9 $\pm$ 5 \\
Sabbagh et al. \cite{Sabbagh} & Centrifuge  & 800 & 10 & $< 4 \times 10^{17}$ & $< 1 \times 10^{18}$  \\
Li et al. \cite{CVDLi} & Centrifuge & 800 & 20 \\
Holmes et al. \cite{Holmes}  &   ${}^{28}\mathrm{Si}^{-}$ ion implanter & 250 & 160 &  $1 \times 10^{17}$ & $3 \times 10^{17}$ \\
Tang et al. \cite{Hypothermal1}  & Magnetic separation  & 0.832 & 0.490 &  $9.5 \times 10^{18}$ & $2.1 \times 10^{18}$ \\
\textbf{This work - sample 1} & \bm{${}^{28}\mathrm{Si}^{++}$} \textbf{FIB} & \textbf{12} \bm{$\pm$} \textbf{2.3} & \textbf{6.0}  \bm{$\pm$} \textbf{1.7}\\
\textbf{This work - sample 2} & \bm{${}^{28}\mathrm{Si}^{++}$} \textbf{FIB}& \textbf{2.3} \bm{$\pm$} \textbf{0.7}& \textbf{0.6}  \bm{$\pm$} \textbf{0.4}\\
\textbf{This work - sample 3} & \bm{${}^{28}\mathrm{Si}^{++}$} \textbf{FIB}& \textbf{6.1} \bm{$\pm$} \textbf{0.9}& \textbf{2.4}  \bm{$\pm$} \textbf{0.6}& \bm{$< 2.5 \times 10^{15}$} & \bm{$< 2.5 \times 10^{15}$} \\
\end{tabular}
\end{ruledtabular}
 \end{adjustbox}
\end{table*}

For prototype silicon based quantum computer devices based on spins it is only necessary to enrich the localised volume that houses the qubits \cite{vandersypen2017interfacing}. Our method enriches localised volumes by employing a focused ${ }^{28}\mathrm{Si}$ ion beam (\textbf{Figure \ref{fig:1}}(a)). This builds on the method employed by Holmes et al. \cite{Holmes} who demonstrated a depletion of ${ }^{29}\mathrm{Si}$ to a level of 250 ppm through the use of a 45 keV ${}^{28}\mathrm{Si}^{-}$ broad ion beam. The focused beam provides a higher beam current intensity leading to relatively short irradiation times that would otherwise be impractical on a broad beam ion implanter. We show that this approach results in a similar residual ${ }^{29}\mathrm{Si}$ and ${ }^{30}\mathrm{Si}$ level as the ``kg-2 project'' but with a method that employs standard natural silicon starting materials.

\section{Method and Results}

\subsection{Focused ion beam enrichment}

\begin{figure}[h!]
\includegraphics[width = \linewidth]{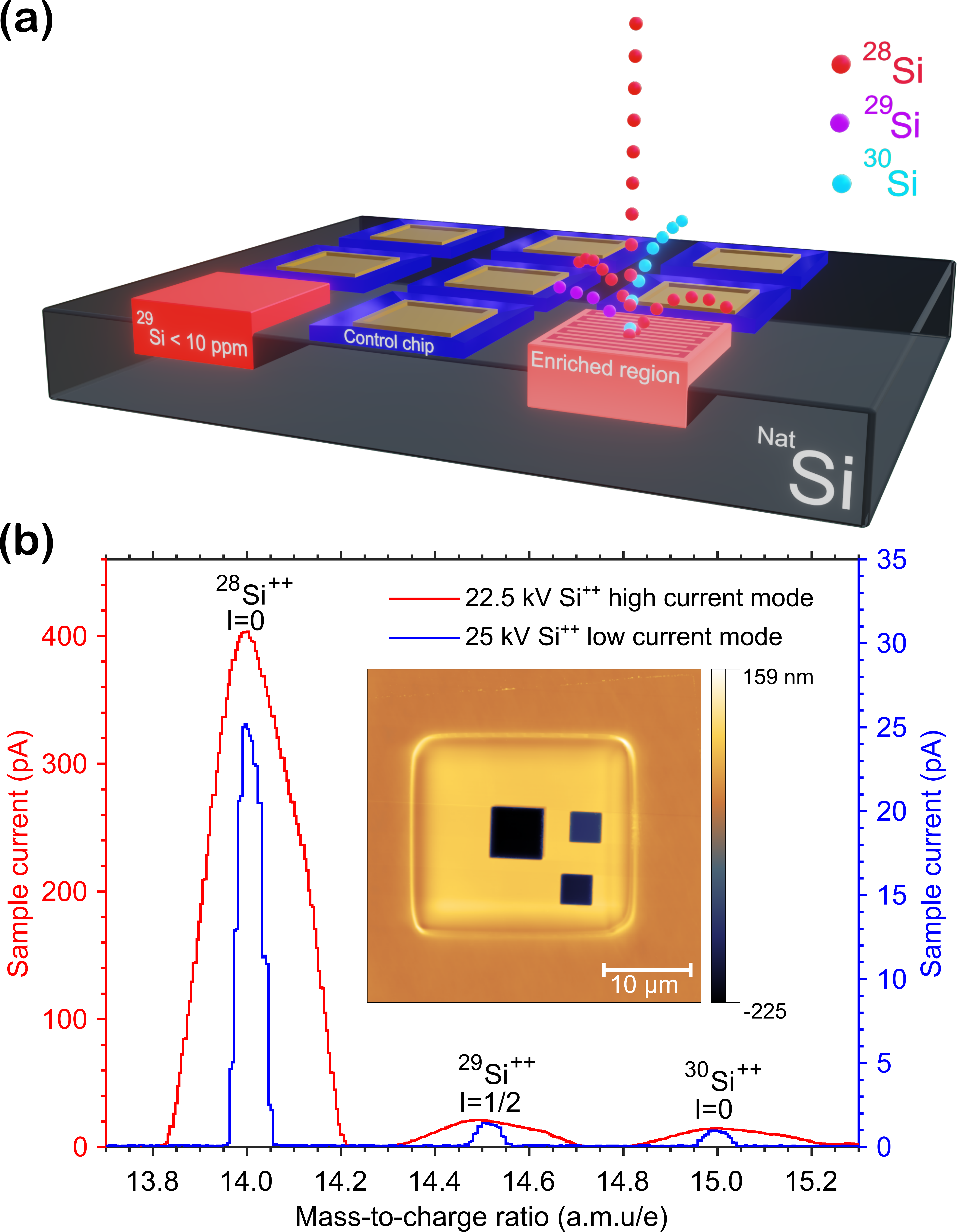} 
\caption{\label{fig:1} Focused ion beam isotopic enrichment using the P-NAME tool. (a) Schematic depicting the isotopic enrichment of localised volumes using a focused ion beam composed of ${ }^{\mathrm{28}}\mathrm{Si}$ where CMOS qubit control electronics are shown to be integrated following the architecture of Vandersypen et al. \cite{vandersypen2017interfacing}. (b) $\mathrm{Si^{++}}$ Wien filter scans highlighting the isotopic mass resolution of  the P-NAME tool. The red high-current scan is taken at an anode voltage of 22.5 kV whereas the blue low-current scan is taken using an anode voltage of 25 kV. The inset of the plot shows the AFM surface map of the as-implanted enriched volume of sample 1 taken after SIMS analysis. The lighter raised square is the implanted area while the dark squares are areas where SIMS has been performed.} 
\end{figure}

Enrichment was performed with a ${}^{28}{\mathrm{Si}}$ ion beam provided by the platform for nanoscale advanced materials engineering (P-NAME) tool developed by Ionoptika Ltd. The P-NAME tool is a focused ion beam (FIB) system that incorporates a liquid metal alloy ion source (LMAIS) from which an ion beam is extracted by field ionization from a sharp emitter tip and subsequently accelerated to keV energies. The P-NAME ion column and sample chamber are held at a base pressure of $10^{-8}$ mbar and $10^{-9}$ mbar, respectively, thus minimising contamination of the sample from residual gas. An \textbf{\textit{E}} $\times$ \textbf{\textit{B}} Wien filter is employed to separate isotopes emitted from the source. Ion optical elements in the column provide beam focusing and scanning, filtration of neutral ions and beam collimation apertures. The sample is placed on a stage angled at 3\textdegree{} with respect to the beam in order to minimise ion channeling. By recording the beam current on the sample stage with a Faraday cup as a function of the applied potential in the Wien filter, it is possible to record and select a specific ion mass from the source as shown in Figure \ref{fig:1}(b). The mass resolution of the Wien filter used here, defined as \textit{M}$/\Delta$\textit{M} for the ${\mathrm{Si}^{++}}$ isotopes, is on average 55 $\pm$ 12. Here the resolution for each isotope is obtained from the Gaussian fitting parameters of the respective Wien filter scan peak.\\ 

  In this work a series of samples were produced with localised enriched volumes implanted with ${ }^{28} \mathrm{Si^{++}}$ fluences within the range $0.84 \, \times \, 10^{19} - 1.18 \times 10^{19}  \,\mathrm{ions}$ 
  $\mathrm{cm}^{-2}$. The LMAISs used were either AuSiSb or AuSiEr eutectic alloys. An accelerating anode potential of 22.5 kV along with the doubly charged ${ }^{28} \mathrm{Si^{++}}$ ions meant that the kinetic energy of the ions composing the beam was equal to 45 keV. This beam energy was predicted to produce a near-planar implanted volume as TRIDYN \cite{MOLLER1984814, MOLLER1988355} modelling has previously shown a sputtering yield of near one at this energy \cite{Holmes}. 
  
For this work an ion beam current of the order of 350 pA was focused into a typical beam diameter of $\sim$ 500 nm which was raster scanned over a 22 $\, \mu \mathrm{m}$ $\times$ 22 $\, \mu \mathrm{m}$ implantation area. For the flip-flop donor qubit architecture with an inter-qubit spacing of 200 nm this implanted area could house a 12,000 flip-flop qubit array \cite{tosi2017silicon}. Sample-specific parameters are provided in \textbf{Table \ref{tab:table1}}. Following implantation a two-step annealing recipe was used to regrow the amorphous volume resulting from the implantation. The first annealing step of 620 $ \degree \mathrm{C}$ for 10 minutes, was followed immediately by a second step of 1000 $ \degree \mathrm{C}$ for 5 s \cite{Holmes}. The ion beam current was measured in a Faraday cup adjacent to the sample before and after the sample irradiation and the  incident beam current inferred from the average of these two measurements, which typically agreed to a precision of better than 4\%.  The beam stability during irradiation was monitored by recording the current from an upstream collimator. The chamber is also equipped with a scanning electron microscope that is used to identify specific locations for ion irradiation with respect to pre-fabricated location markers.

\begin{table*}
 \caption{\label{tab:table1} A summary of the implantation and annealing parameters used for the fabrication of the isotopically enriched ${ }^{28} \mathrm{Si}$ samples outlined in this work. Each sample was composed of a single localised 22 $\, \mu \mathrm{m}$ $\times$ 22 $\mu$m square enriched area and enriched using a ${ }^{28} \mathrm{Si^{++}}$ focused ion beam accelerated under a potential of 22.5 kV to yield a kinetic energy of 45 keV. Samples 3 and 5 were annealed under an ambient Ar atmosphere using a two-step process
 consisting of an initial step of 620  $\degree\mathrm{C}$ for 10 minutes followed by a second step of 1000 $ \degree \mathrm{C}$ for 5 s. In instances where significant ion beam astigmatism was present two ion beam width values measured along orthogonal axes are quoted in order to quantify the extent of the beam astigmatism.}
\centering
\begin{adjustbox}{width=1\textwidth}
\begin{tabular}{ccccc}
    \hline  \noalign{\vskip 0.03in} 
    Sample & ${ }^{28} \mathrm{Si^{++}}$ ion fluence [$\times 10^{19} \, \mathrm{ions} \, \mathrm{cm}^{-2}$] & Ion beam current [pA] & Ion beam width [nm] & Annealed \\
    \hline
 
    1 & $(0.84 \pm 0.01)$   & 361 $\pm$ 3 & 730 $\pm$ 60 & No \\ 
    2 & $(1.18 \pm 0.04)$  & 380 $\pm$ 14 & 730 $\pm$ 60 & No \\ 
    3 & $(1.13 \pm 0.01)$  & 340 $\pm$ 8 & 400 $\pm$ 60/ 230 $\pm$ 30 & Yes\\ 
    4 & $(1.08 \pm 0.01)$   & 330 $\pm$ 2  & 480 $\pm$ 6/ 310 $\pm$ 30 & No  \\ 
    5 & $(0.84 \pm 0.08)$ & 330 $\pm$ 2 ($\sim$ 53 $\%$) +  159 $\pm$ 2 ($\sim$ 47 $\%$)  & 320 $\pm$ 70 & Yes \\ 

    \hline
  \end{tabular}
  \end{adjustbox}
\end{table*}

\begin{figure*}
\centering
\includegraphics[width = 0.95\linewidth]{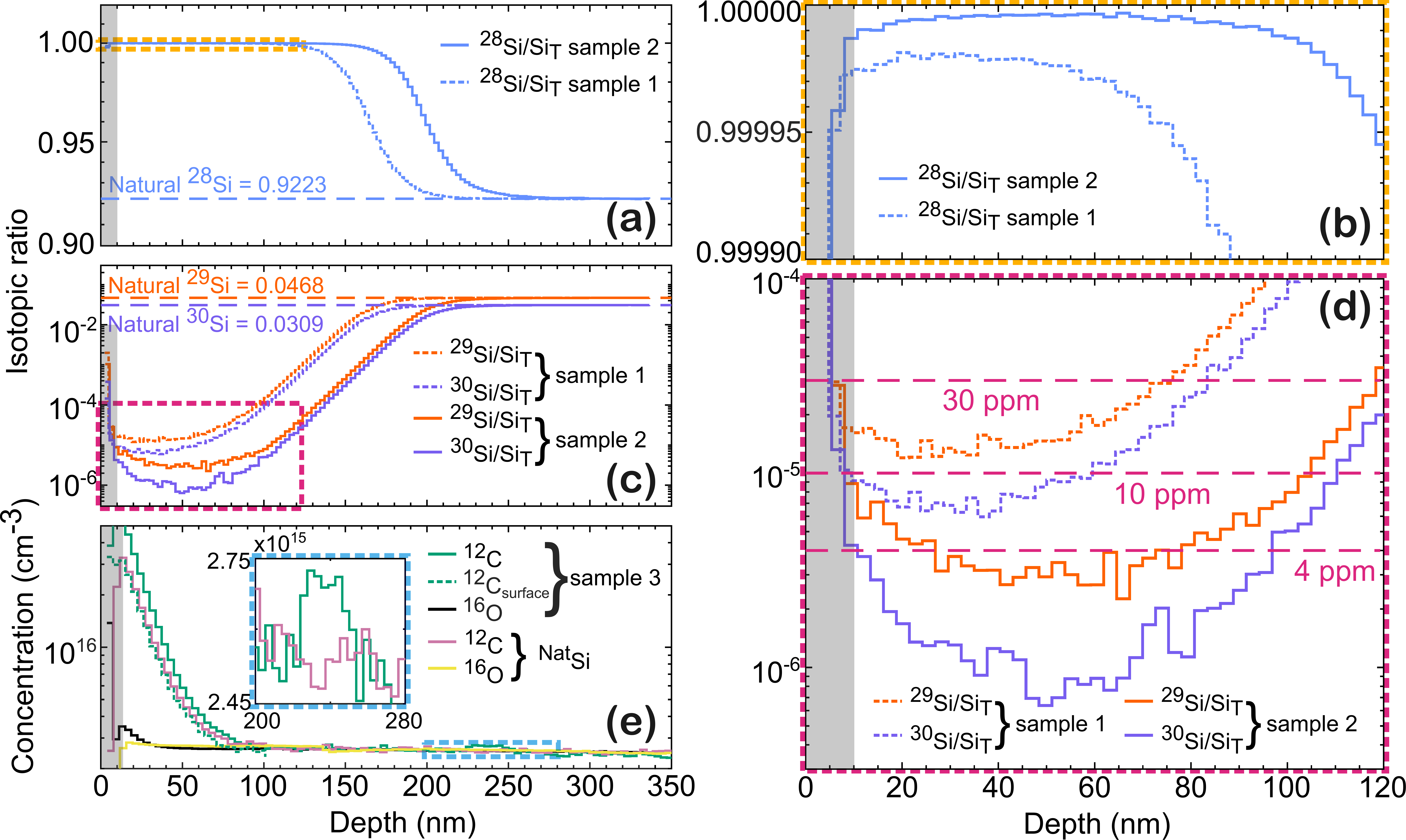}
\caption{\label{fig:Combined} NanoSIMS isotopic elemental analysis on isotopically enriched samples. (a) - (d) show depth profiles of the measured isotopic ratios for each of the three silicon isotopes from samples 1 (dotted line) and 2 (solid line), which have been normalised by dividing each by $\mathrm{Si}_{\mathrm{T}}$ = ${ }^{28} \mathrm{Si} + { }^{29} \mathrm{Si}+{ }^{30} \mathrm{Si}$. (b) and (d) are enlarged from the dotted regions outlined within (a) and (c), respectively. (e) shows the ${ }^{12} \mathrm{C}$ and ${ }^{16} \mathrm{O}$ concentration depth profiles for sample 3 which was annealed following enrichment. The near-surface carbon signal shown by the dotted green line in part (e) was selected such that surface contaminants introduced during sample transfer were excluded (supplementary information Figure S5). This near-surface signal along with the oxygen profile show the limited introduction of impurities within the enriched region relative to the unimplanted substrate. The figure inset to (e) illustrates ${}^{12}\mathrm{C}$ gettering observed near the bottom of the enriched volume where the C concentration rises above the minimum detectable limit. The shaded grey box in each plot indicates the portion of the SIMS depth profile affected by the ``steady-state'' artefact and thus excluded from the analysis. }
\end{figure*}

\subsection{Enrichment measurement}

In order to quantify the enrichment and residual contamination level within the implanted regions nano-scale secondary ion mass spectrometry (NanoSIMS) analysis, using a $\mathrm{Cs}^{+}$ beam to sputter the sample, was performed on samples 1, 2 and 3. Depth profiles of the measured isotopic ratios for each of the three silicon isotopes, normalised by dividing each by the sum of the three silicon isotope count values ($\mathrm{Si}_{\mathrm{T}}$ = ${ }^{28} \mathrm{Si} + { }^{29} \mathrm{Si}+{ }^{30} \mathrm{Si}$), are shown for the two as-implanted samples 1 and 2 in \textbf{Figure \ref{fig:Combined}}.  The initial SIMS signal (shown within the shaded grey areas of each profile in Figure \ref{fig:Combined}) is influenced by the well-known ``steady-state''  artefact whereby a threshold dose of $\mathrm{Cs}^{+}$ has to be implanted before both the $\mathrm{Cs}^{+}$ implantation and sample sputtering rate reach a dynamic equilibrium \cite{NanoSIMS} and can therefore be excluded from the true profiles. For the extreme fluences employed here the  variation in the Si isotopic ratio profiles of all samples indicates that atom mixing occurs throughout the depth of the enriched volume by a combination of mechanisms including deposition, surface sputtering and forward recoils.\\ 

Atomic force microscopy (AFM) measurements on as-implanted samples for fluences above $\sim 1 \times 10^{18} \,  \mathrm{ions} \, \mathrm{cm}^{-2}$ show a consistent increase in the volume of the implanted area that scales with the implanted ion fluence (supplementary information Figure S1). This effect is not accounted for in TRIDYN simulations which predict a sputtering yield of one. We note that the experimental deviation from the simulations for this higher fluence regime likely arise from ion interactions not accounted for in the modelling.  An AFM surface map of sample 1 post-SIMS analysis, inset to Figure \ref{fig:1}(b), shows the presence of both this increased volume that is apparent from surface swelling and also a prominent ridge around the boundary of the implanted area. UV Raman spectra taken from this boundary ridge on an as-implanted sample (supplementary information Figure S2) identifies this as being composed of $\mathrm{sp}^{2}$ amorphous carbon. Further Raman spectra show the absence of carbon from both the unimplanted surrounding substrate and within the implanted area. The accumulation of $\mathrm{sp}^{2}$ amorphous carbon has been observed from MeV $\mathrm{He}^{+}$ irradiated Cu surfaces previously exposed to the ambient atmosphere \cite{KELLOCK1997742}, where the mechanism was assumed to be ion beam induced dissociation of a hydrocarbon absorbate layer.\\  
\begin{figure*}
\centering
        \includegraphics[width = 0.75\linewidth]{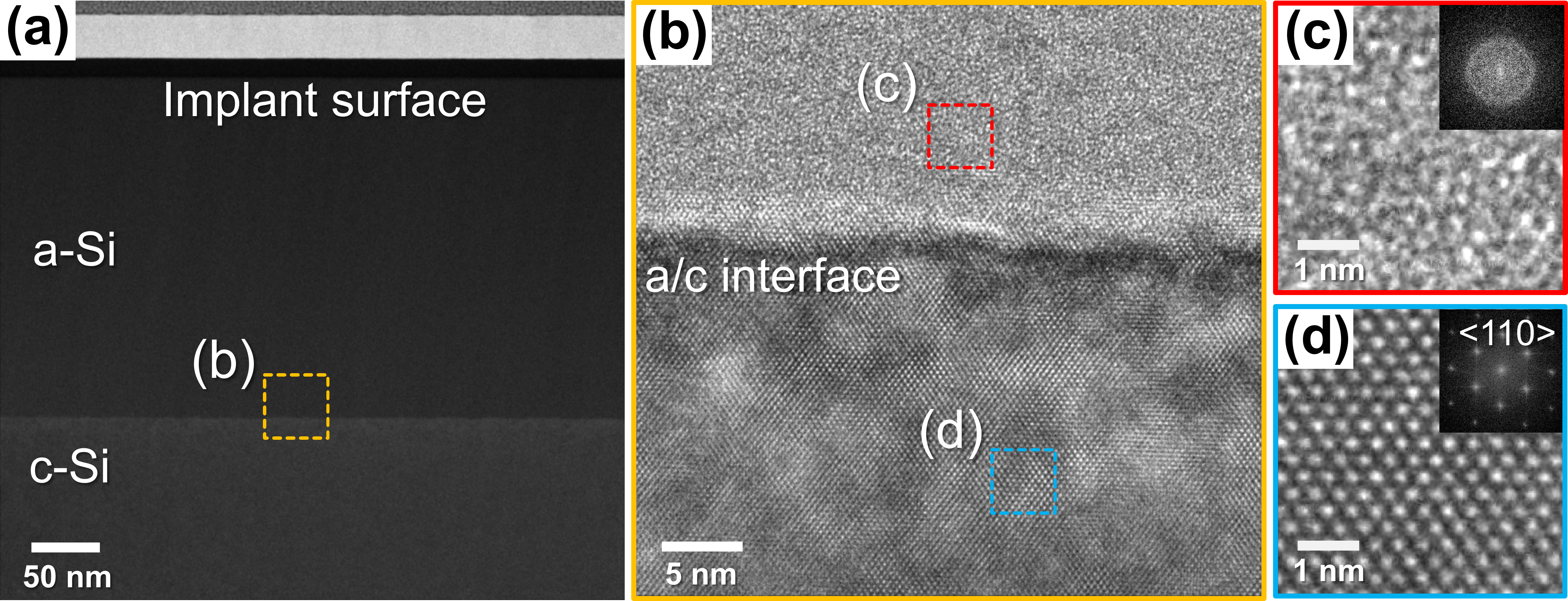}  
            \caption{\label{fig:TEM_Amorphous} Cross-sectional TEM images of the as-implanted sample 4 showing the amorphous silicon (a-Si) implanted volume and the original crystalline silicon (c-Si) below. (a) shows a HAADF STEM cross-sectional image of the sample, the amorphous-crystalline (a/c) interface is shown to be very sharp and $\sim$ 250 nm below the surface. (b) shows a high-resolution cross-sectional TEM image along the amorphous-crystalline interface. (c) and (d) show  the HRTEM images taken from within the amorphous (enriched) region (red area in (b)) and the crystalline substrate beneath (blue area in (b)), respectively. The associated FFT taken from each image is shown in the inset. The rings observed in the FFT of part (c) taken above the interface indicate that the implanted volume is amorphous as compared to the single crystalline substrate beneath.}
    \end{figure*}

NanoSIMS analysis of the enriched volume following the solid phase epitaxial (SPE) regrowth of the implanted volume of sample 3 is shown in Figure \ref{fig:Combined}(e), where the ${  }^{12} \mathrm{C}$ and ${  }^{16} \mathrm{O}$ NanoSIMS depth profiles taken from both within the enriched volume  as well as the unimplanted surrounding ${ }^{\mathrm{Nat}}\mathrm{Si}$ are observed to be comparable.
 Concentration values were determined by assuming the upper limit for the original wafer background C and O concentration of $2.5 \times 10^{15} \, \mathrm{cm}^{-3}$ provided by the wafer supplier. The slight increase in the sample O signal near the surface is attributed to surface contaminants. The C signal taken from the sample also shows evidence of gettering at the approximate depth of the original amorphous-crystalline interface as a result of the epitaxial regrowth process. UV Raman spectra (supplementary information Figure S3) also  confirm the SPE regrowth of the enriched volume after annealing. The minimum concentrations of ${}^{29}\mathrm{Si}$ and  ${}^{30}\mathrm{Si}$ within the enriched volumes of sample 1-3 are shown in Table \ref{tab:tablesummary}, where the enrichment level for all three samples is shown to be comparable.

\subsection{Structural characterisation }

\begin{figure}
            \includegraphics[width = \linewidth]{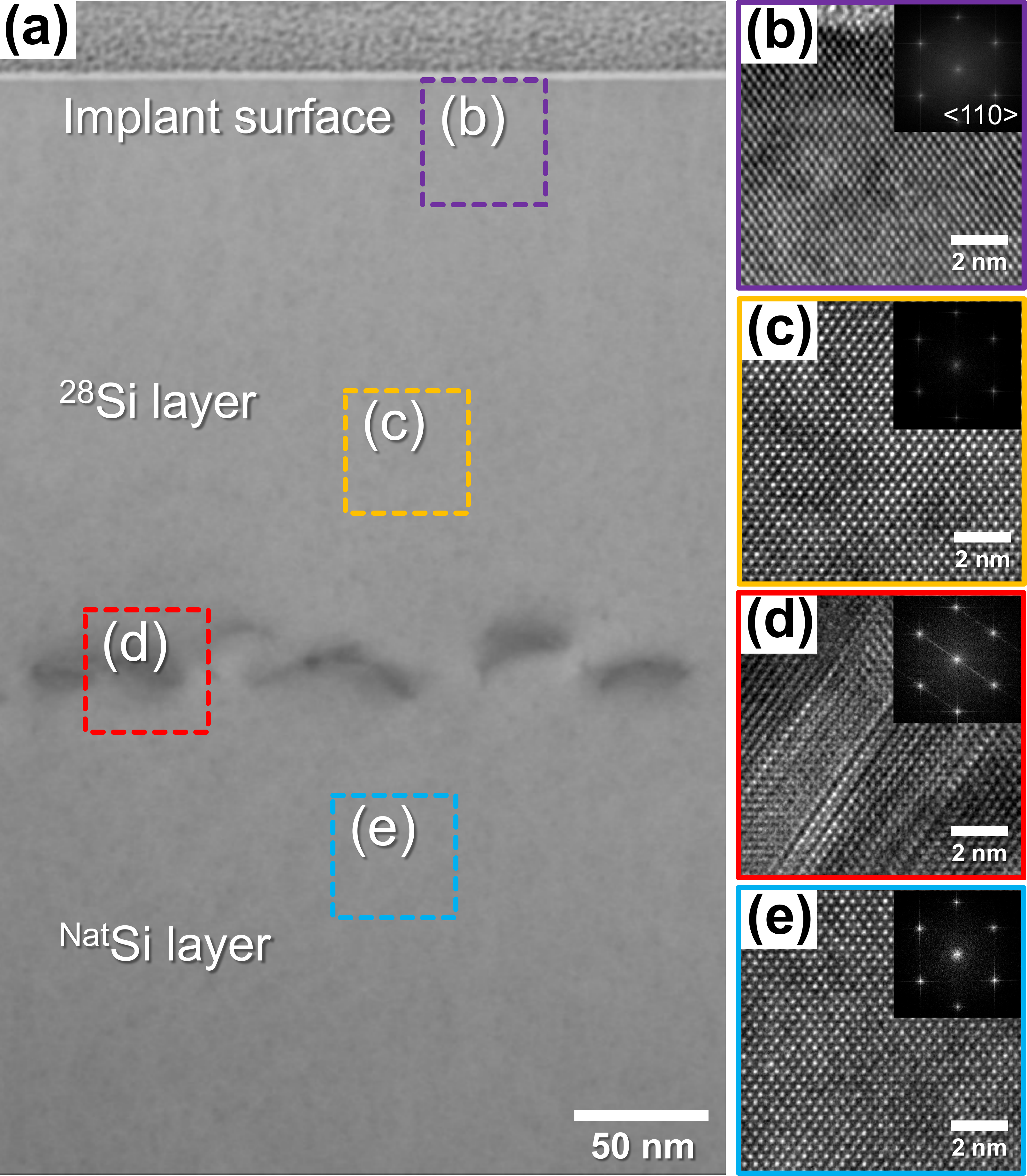}            \caption{\label{fig:TEM_Annealed} Cross-sectional TEM images of sample 5 post-anneal. (a) shows the bright field (BF) STEM cross-sectional image of the annealed sample. The enriched volume is shown to be free of any defects barring a layer found $\sim$ 225 nm below the surface. A series of HRTEM images taken of areas across the sample corresponding to different distances from the implanted surface (as denoted by the coloured boxes in (a)) are shown in (b) - (e) where the associated FFT taken from each image is shown in the inset. The similarity of the FFTs taken from the near-surface region (b) and the single crystalline substrate beneath the implanted volume (e) verifies the complete single crystalline regrowth of the enriched volume. (d) shows the presence of an extrinsic dislocation loop within the defect layer. 
            }
    \end{figure}

Transmission electron microscopy (TEM) analysis was carried out on the as-implanted enriched volume within sample 4 in order to determine the thickness of the  amorphous volume above the single crystal substrate prior to annealing. \textbf{Figure \ref{fig:TEM_Amorphous}} shows the cross-sectional TEM images and corresponding fast fourier transforms (FFTs) taken from sample 4. From the high-angle annular dark-field (HAADF) image shown in Figure \ref{fig:TEM_Amorphous}(a) the amorphous volume is observed to be $\sim$ 250 nm deep which is comparable to the NanoSIMS depletion profile of sample 2 (Figure \ref{fig:Combined}) which shows a $\sim$ 270 nm deep enriched volume where a fluence variation accounts for the slight discrepancy between these samples. The diffuse rings present in the FFT taken from the high-resolution TEM image (HRTEM) shown Figure \ref{fig:TEM_Amorphous}(c) shows the amorphous nature of the implanted volume. Following the enrichment irradiation, a SPE anneal regrows the amorphous volume to single crystal from the substrate. The crystalline phase of the sub-surface enriched volume of sample 5 after annealing was determined through subsequent TEM analysis, shown in \textbf{Figure \ref{fig:TEM_Annealed}}. The bright field scanning TEM (BF-STEM) image shown in Figure \ref{fig:TEM_Annealed}(a) shows that the enriched volume is free of defects except for a clearly defined defect layer situated $\sim$ 225 nm from the surface. These defects are formed below the amorphous-crystalline interface upon annealing the amorphous volume \cite{OswaldRipening, jones1988systematic} whereby the interstitial abundant region below the interface leads to the formation of extrinsic dislocation loops. The HRTEM image taken from this defect layer in Figure \ref{fig:TEM_Annealed}(d) shows such a dislocation loop. The near-surface HRTEM image and corresponding FFT shown in Figure \ref{fig:TEM_Annealed}(b) show that the near-surface region is single crystal demonstrating the SPE regrowth of the amorphous volume up to the surface. For most silicon based donor or dot architectures the qubits are located in the top $\sim$ 20 nm  of the substrate which is far above the buried residual defects. For example the flip-flop donor qubit architecture employs ${}^{31}{\mathrm{P}}$ donors implanted $\sim$ 15 nm below a thin surface oxide \cite{tosi2017silicon}.

\section{Conclusions}

We have demonstrated the use of a focused beam of ${ }^{28} \mathrm{Si^{++}}$ ions to produce highly isotopically enriched volumes embedded in the surface of natural silicon substrates. NanoSIMS analysis has shown an isotopically enriched volume produced with a nominal ${ }^{28} \mathrm{Si^{++}}$ fluence of $(1.18 \, \pm \, 0.04) \times 10^{19} \, \mathrm{ions} \, \mathrm{cm}^{-2}$ has a  residual ${ }^{29} \mathrm{Si}$ level at a minimum of 2.3 ppm. Further NanoSIMS analysis on an annealed single crystalline enriched sample shows negligible introduction of impurities into the enriched volume produced as a result of the combined annealing and implantation process. This indicates that the contamination level of the enriched volume is largely dependent on the chemical purity of the starting ${ }^{\mathrm{Nat}} \mathrm{Si}$ wafer. TEM analysis verifies the amorphous nature of the as-implanted volume as well as the single crystalline SPE regrowth upon annealing. Enriched single crystal silicon produced with this method has both high chemical and isotopic purity. A near-term application of our technique will be to fabricate an improved flip-flop qubit device.  Previous work reported a single ${ }^{31} \mathrm{P}$ donor flip-flop qubit device \cite{Savytskyy} constructed in an epitaxial enriched ${ }^{28} \mathrm{Si}$ epilayer in which the ${ }^{29} \mathrm{Si}$ concentration had been depleted to 730 ppm.  At this level of depletion $\sim$ 10 ${ }^{29} \mathrm{Si}$ atoms could be located within the Bohr radius of the ${ }^{31} \mathrm{P}$ donor \cite{Freezing} some of which may perturb the qubit.   In fact, by use of NMR RF pulses tuned to the ${ }^{29} \mathrm{Si}$ nuclear transitions, the signal from at least 3 ${ }^{29} \mathrm{Si}$ nuclei were observed by their effect on the flip-flop qubit.  The most highly depleted material reported in the present work, with two orders of magnitude better depletion, could be used to construct a large-scale flip-flop qubit device, also incorporating direct-write depleted vias for qubit coupling, that is largely free from such perturbations. The unique ability of the FIB to direct write localised enriched volumes, together with the capability to implant the heavy group-V donor ${ }^{123} \mathrm{Sb}$ (\textit{I} = 7/2) without breaking vacuum through the use of the AuSiSb LMAIS, shows the longer term potential of the FIB enrichment method for use in silicon based quantum computation architectures incorporating high spin nuclei \cite{asaad2020coherent}. The high enrichment level reported here will open new avenues in probing fundamental qubit interactions within entangled heavy-donor multi-qubit devices in a regime where the coupling to the residual host nuclear spin bath is orders of magnitude less significant than at present.


\section*{Experimental Section}
\subsection*{Sample fabrication}
 The un-implanted sample substrate was an undoped $ >10 \, \mathrm{k} \Omega \cdot \mathrm{cm} $  [100] 500 $\mathrm{\mu m}$ ${ }^{\mathrm{Nat}} \mathrm{Si}$ wafer. The background levels of carbon and oxygen were both quoted by the supplier (University Wafer, Inc.) as having an upper limit of $2.5 \times 10^{15} \, \mathrm{cm}^{-3}$. All samples were cleaned and then dipped in hydroflouric acid (HF) prior to loading into the P-NAME sample vacuum chamber in order to remove the native silicon oxide. The ion beam current was measured using a Faraday cup on the sample stage which was connected to a Keithley 6285 picoammeter. A 300 $\mu$m resolving aperture was in place along the ion column for the enrichment work presented. The mass resolution of the Wien filter used for the enrichment was 55 $\pm$ 12  \textit{M}$/\Delta$\textit{M} which ensured a high isotopic purity within the beam as demonstrated by the present results. The nominal ion beam profile on the sample was assumed to be a Gaussian point spread function where $2\sigma$ was taken as the ion beam width. The ion beam width was measured using the 20/80 knife-edge method from a line-scan taken over a sharp high-contrast metal edge using the secondary electron signal.  The signal intensity as a function of the beam position in the line scan was fitted with an error function using the ``GaussFit'' plugin (based on the ASTM E986-9731) within the ImageJ software package and the 20\% to 80\% amplitude was taken as the ion beam width. Following irradiation some samples were annealed in an ambient Ar atmosphere using an Annealsys model AS-One 100 rapid thermal annealer.
 
\subsection*{Atomic force microscopy}  AFM measurements for sample 3 were carried out using a Nanosurf coreAFM 190Al-G tip which was operated in tapping mode. AFM measurements for samples 1 and 2 were taken using a neaSNOM scattering-type near-field optical microscope (s-SNOM) operated in tapping-mode using a Pt-coated, grounded AFM tip. For all the samples the raw AFM image height correction was carried out using the Gwyddion software package (version 2.62) to measure the swelling of the implanted volume.

\subsection*{NanoSIMS analysis} A CAMECA NanoSIMS 50L instrument employing a 16 keV $\mathrm {Cs}^{+}$ beam with a beam current of 0.43 - 0.48 pA was used to sputter the target sample in turn generating negative secondary ions from the surface. A double focusing mass spectrometer was employed to detect the ions produced. The instrument is equipped with seven detectors that each register a specific pre-defined ion species. For both samples 2 and 3 the detector configuration was ${  }^{12} \mathrm{C}^{-}$, ${  }^{16} \mathrm{O}^{-}$,(${  }^{12} \mathrm{C}$${  }^{14} \mathrm{N})^{-}$, ${  }^{28} \mathrm{Si}^{-}$, ${  }^{29} \mathrm{Si}^{-}$, ${  }^{30} \mathrm{Si}^{-}$, ${  }^{16} \mathrm{O}^{-}_{2}$ whereas for sample 1 the configuration was ${  }^{12} \mathrm{C}^{-}$, ${  }^{28} \mathrm{Si}^{-}$, ${  }^{29} \mathrm{Si}^{-}$, ${  }^{30} \mathrm{Si}^{-}$, ${  }^{16} \mathrm{O}^{-}_{2}$, ${  }^{12} \mathrm{C}^{-}_{3}$. The mass resolution of the instrument was set to be greater than 6000  \textit{M}$/\Delta$\textit{M} so that the ${ }^{29} \mathrm{Si}$ could be separated from the ${ }^{28} \mathrm{SiH}$ mass interference and the ${ }^{30} \mathrm{Si}$ from the ${ }^{29} \mathrm{SiH}$. 5 $\mathrm{\mu}$m by 5 $\mathrm{\mu}$m raster scans were taken from within the center of each of the enriched volumes except for sample 3 where a series of raster scans were taken from different areas across the enriched volume. Each scan was composed of 256 by 256 pixels leading to a pixel resolution of 19.5 nm per pixel with a dwell time set to 2 ms per pixel. Slit positions, designated D1-4 (150 $\mu$m in diameter), ES-3 (30 $\mu$m width) and AS-2 (200 $\mu$m width) were utilised. The instrument analysis chamber was held at a vacuum of the order of $10^{-10}$ mbar. All samples analysed using the tool were HF dipped prior to loading into the SIMS machine chamber. NanoSIMS data were processed using either the ``Open MIMS image'' plugin found in the ImageJ software package or the L'image software package developed by L. Nittler at the Carnegie Institution of Washington.  AFM measurements of the sputtered SIMS craters were taken to calibrate the depth of the isotope profiles after the SIMS analysis assuming a constant sputtering rate.


\subsection*{TEM analysis} TEM analysis was carried out on both sample 4 and 5 in order to verify the amorphous implanted volume depth and the regrowth quality of the enriched volume upon annealing, respectively. Cross-sectional lamella taken from both samples were prepared by FIB milling using an FEI Helios 660 FIB. Prior to lamella FIB extraction both samples were ex situ coated with a thin carbon layer followed by a AuPd layer to mitigate charging effects and to protect the surface from FIB damage. A further approximately 300 nm thick sacrificial Pt layer was then deposited via electron beam to protect the film from any Ga ion implantation during the milling process. This was followed by a final $\sim$ 3 $\mathrm{\mu}$m thick Pt protective layer that was deposited using the Ga ion beam. Cross-sectional TEM and HAADF images were taken of both samples using a Talos F200X S/TEM tool operating at 200keV. The TEM image shown in Figure \ref{fig:TEM_Annealed}(a) was filtered in order to reduce the visibility of surface damage introduced during the lamellae sample preparation (raw data in supplementary information Figure S7).

\medskip
\textbf{Acknowledgements}  
This work was funded by the Australian Research Council Discovery Project DP220103467 and the University of Melbourne/ University of Manchester collaboration scheme.  D.N.J acknowledges the support of a Royal Society (UK) Wolfson Visiting Fellowship RSWVF/211016. R.A acknowledges the support of a Melbourne Research Scholarship.
This work was funded by the EPSRC grants EP/R025576/1, EP/V001914/1 and EP/R00661X/1 (Henry Royce Institute) and by capital investment by the University of Manchester. S.J.H acknowledges the European Research Council under the European Union's Horizon 2020 research and innovation program (grant agreement No. 715502, EvoluTEM). Electron microscopy access was supported by the Henry Royce Institute for Advanced Materials, funded through EPSRC grants EP/S01936 7/1, EP/P025021/1 and EP/P025498/1. The NanoSI-MS was funded by UK Research Partnership Investment Funding (UKRPIF) Manchester RPIF Round 2 and also supported by the Henry Royce Institute for Advanced Materials. B.G and J.L.B thank EPSRC for support through grants EP/T01914X/1, EP/S037438/1 and capital investment via Univeristy of Manchester and Royce Institute. J.L.B also acknowledges funding from UKRI via MR/T022140/1. The authors acknowledge useful discussions with J.C. McCallum, A.M. Jakob, S.Q. Lim and A. Morello.

\medskip
\bibliographystyle{MSP}
\bibliography{apssamp}

\begin{thebibliography}{10}
\providecommand{\url}[1]{\texttt{#1}}
\providecommand{\urlprefix}{URL }

\bibitem{General_adv_mat}
E.~H. Lock, J.~Lee, D.~S. Choi, R.~G. Bedford, S.~P. Karna, A.~K. Roy,
\newblock \emph{Advanced Materials} \textbf{2023}, \emph{35}, 27 2201064.

\bibitem{takeda2022quantum}
K.~Takeda, A.~Noiri, T.~Nakajima, T.~Kobayashi, S.~Tarucha,
\newblock \emph{Nature} \textbf{2022}, \emph{608}, 7924 682.

\bibitem{saraiva2022materials}
A.~Saraiva, W.~H. Lim, C.~H. Yang, C.~C. Escott, A.~Laucht, A.~S. Dzurak,
\newblock \emph{Advanced Functional Materials} \textbf{2022}, \emph{32}, 3
  2105488.

\bibitem{kane1998Silicon}
B.~E. Kane,
\newblock \emph{Nature} \textbf{1998}, \emph{393}, 6681 133.

\bibitem{Semiconductorqubitsinpractice}
A.~Chatterjee, P.~Stevenson, S.~De~Franceschi, A.~Morello, N.~P. de~Leon,
  F.~Kuemmeth,
\newblock \emph{Nature Reviews Physics} \textbf{2021}, \emph{3}, 3 157.

\bibitem{muhonen2014storing}
J.~T. Muhonen, J.~P. Dehollain, A.~Laucht, F.~E. Hudson, R.~Kalra,
  T.~Sekiguchi, K.~M. Itoh, D.~N. Jamieson, J.~C. McCallum, A.~S. Dzurak,
  A.~Morello,
\newblock \emph{Nature nanotechnology} \textbf{2014}, \emph{9}, 12 986.

\bibitem{dehollain2016optimization}
J.~P. Dehollain, J.~T. Muhonen, R.~Blume-Kohout, K.~M. Rudinger, J.~K. Gamble,
  E.~Nielsen, A.~Laucht, S.~Simmons, R.~Kalra, A.~S. Dzurak,
\newblock \emph{New Journal of Physics} \textbf{2016}, \emph{18}, 10 103018.

\bibitem{CMOScompatbilitySi}
M.~Khoury, M.~Abbarchi,
\newblock \emph{Journal of Applied Physics} \textbf{2022}, \emph{131}, 20
  200901.

\bibitem{29Spin}
E.~Abe, A.~M. Tyryshkin, S.~Tojo, J.~J.~L. Morton, W.~M. Witzel, A.~Fujimoto,
  J.~W. Ager, E.~E. Haller, J.~Isoya, S.~A. Lyon, M.~L.~W. Thewalt, K.~M. Itoh,
\newblock \emph{Phys. Rev. B} \textbf{2010}, \emph{82} 121201.

\bibitem{Abundances}
K.~J.~R. Rosman, P.~D.~P. Taylor,
\newblock \emph{Journal of Physical and Chemical Reference Data} \textbf{1998},
  \emph{27}, 6 1275.

\bibitem{Sousa2003}
R.~de~Sousa, S.~D. Sarma,
\newblock \emph{Physical Review B - Condensed Matter and Materials Physics}
  \textbf{2003}, \emph{68} 1153221.

\bibitem{Witzel2006}
W.~M. Witzel, S.~D. Sarma,
\newblock \emph{Physical Review B - Condensed Matter and Materials Physics}
  \textbf{2006}, \emph{74}.

\bibitem{Witzel2010}
W.~M. Witzel, M.~S. Carroll, A.~Morello, Łukasz Cywiński, S.~D. Sarma,
\newblock \emph{Physical Review Letters} \textbf{2010}, \emph{105}.

\bibitem{itoh_watanabe_2014}
K.~M. Itoh, H.~Watanabe,
\newblock \emph{MRS Communications} \textbf{2014}, \emph{4}, 4 143–157.

\bibitem{Avogadro}
P.~Becker, H.-J. Pohl, H.~Riemann, N.~Abrosimov,
\newblock \emph{Physica status solidi (a)} \textbf{2010}, \emph{207}, 1 49.

\bibitem{kg-2}
N.~V. Abrosimov, D.~G. Aref’ev, P.~Becker, H.~Bettin, A.~D. Bulanov,
  M.~Churbanov, S.~Filimonov, V.~Gavva, O.~N. Godisov, A.~V. Gusev, T.~V.
  Kotereva, D.~Nietzold, M.~Peters, A.~M. Potapov, H.-J. Pohl, A.~Pramann,
  H.~Riemann, P.-T. Scheel, R.~Stosch, S.~Wundrack, S.~Zakel,
\newblock \emph{Metrologia} \textbf{2017}, \emph{54}, 4 599.

\bibitem{dwyer2014enriching}
K.~J. Dwyer, J.~M. Pomeroy, D.~S. Simons, K.~L. Steffens, J.~W. Lau,
\newblock \emph{Journal of Physics D: Applied Physics} \textbf{2014},
  \emph{47}, 34 345105.

\bibitem{Hypothermal1}
K.~Tang, H.~S. Kim, A.~N.~R. Ramanayaka, D.~S. Simons, J.~M. Pomeroy,
\newblock \emph{Review of Scientific Instruments} \textbf{2019}, \emph{90}, 8
  083308.

\bibitem{tang2020targeted}
K.~Tang, H.~S. Kim, A.~N. Ramanayaka, D.~S. Simons, J.~M. Pomeroy,
\newblock \emph{Journal of physics communications} \textbf{2020}, \emph{4}, 3
  035006.

\bibitem{CVDMAZZOCCHI20191}
V.~Mazzocchi, P.~G. Sennikov, A.~D. Bulanov, M.~F. Churbanov, B.~Bertrand,
  L.~Hutin, J.~P. Barnes, M.~N. Drozdov, J.~M. Hartmann, M.~Sanquer,
\newblock \emph{Journal of Crystal Growth} \textbf{2019}, \emph{509} 1.

\bibitem{Sabbagh}
D.~Sabbagh, N.~Thomas, J.~Torres, R.~Pillarisetty, P.~Amin, H.~George,
  K.~Singh, A.~Budrevich, M.~Robinson, D.~Merrill, L.~Ross, J.~Roberts,
  L.~Lampert, L.~Massa, S.~V. Amitonov, J.~M. Boter, G.~Droulers, H.~G.~J.
  Eenink, M.~van Hezel, D.~Donelson, M.~Veldhorst, L.~M.~K. Vandersypen, J.~S.
  Clarke, G.~Scappucci,
\newblock \emph{Phys. Rev. Appl.} \textbf{2019}, \emph{12} 014013.

\bibitem{CVDLi}
J.-Y. Li, C.-T. Huang, L.~P. Rokhinson, J.~C. Sturm,
\newblock \emph{Applied Physics Letters} \textbf{2013}, \emph{103}, 16 162105.

\bibitem{Holmes}
D.~Holmes, B.~C. Johnson, C.~Chua, B.~Voisin, S.~Kocsis, S.~Rubanov, S.~G.
  Robson, J.~C. McCallum, D.~R. McCamey, S.~Rogge, D.~N. Jamieson,
\newblock \emph{Phys. Rev. Mater.} \textbf{2021}, \emph{5} 014601.

\bibitem{vandersypen2017interfacing}
L.~Vandersypen, H.~Bluhm, J.~Clarke, A.~Dzurak, R.~Ishihara, A.~Morello,
  D.~Reilly, L.~Schreiber, M.~Veldhorst,
\newblock \emph{npj Quantum Information} \textbf{2017}, \emph{3}, 1 34.

\bibitem{MOLLER1984814}
W.~Möller, W.~Eckstein,
\newblock \emph{Nuclear Instruments and Methods in Physics Research Section B:
  Beam Interactions with Materials and Atoms} \textbf{1984}, \emph{2}, 1 814.

\bibitem{MOLLER1988355}
W.~Möller, W.~Eckstein, J.~P. Biersack,
\newblock \emph{Computer Physics Communications} \textbf{1988}, \emph{51}, 3
  355.

\bibitem{tosi2017silicon}
G.~Tosi, F.~A. Mohiyaddin, V.~Schmitt, S.~Tenberg, R.~Rahman, G.~Klimeck,
  A.~Morello,
\newblock \emph{Nature communications} \textbf{2017}, \emph{8}, 1 450.

\bibitem{NanoSIMS}
Y.~Aboura, K.~Moore,
\newblock \emph{Applied Surface Science} \textbf{2021}, \emph{557} 149736.

\bibitem{KELLOCK1997742}
A.~J. Kellock, M.~H. Tabacniks, J.~E.~E. Baglin, N.~S. Somcio, T.~T. Bardin,
  D.~C. Miller,
\newblock \emph{Nuclear Instruments and Methods in Physics Research Section B:
  Beam Interactions with Materials and Atoms} \textbf{1997}, \emph{127-128}
  742.

\bibitem{OswaldRipening}
C.~Bonafos, D.~Mathiot, A.~Claverie,
\newblock \emph{Journal of Applied Physics} \textbf{1998}, \emph{83}, 6 3008.

\bibitem{jones1988systematic}
K.~S. Jones, S.~Prussin, E.~R. Weber,
\newblock \emph{Applied Physics A} \textbf{1988}, \emph{45} 1.

\bibitem{Savytskyy}
R.~Savytskyy, T.~Botzem, I.~F. de~Fuentes, B.~Joecker, J.~J. Pla, F.~E. Hudson,
  K.~M. Itoh, A.~M. Jakob, B.~C. Johnson, D.~N. Jamieson, A.~S. Dzurak,
  A.~Morello,
\newblock \emph{Science Advances} \textbf{2023}, \emph{9}, 6 eadd9408.

\bibitem{Freezing}
M.~T. M\k{a}dzik, T.~D. Ladd, F.~E. Hudson, K.~M. Itoh, A.~M. Jakob, B.~C.
  Johnson, J.~C. McCallum, D.~N. Jamieson, A.~S. Dzurak, A.~Laucht, A.~Morello,
\newblock \emph{Science Advances} \textbf{2020}, \emph{6}, 27 eaba3442.

\bibitem{asaad2020coherent}
S.~Asaad, V.~Mourik, B.~Joecker, M.~A.~I. Johnson, A.~D. Baczewski, H.~R.
  Firgau, M.~T. M{\k{a}}dzik, V.~Schmitt, J.~J. Pla, F.~E. Hudson, K.~M. Itoh,
  J.~C. McCallum, A.~S. Dzurak, A.~Laucht, A.~Morello,
\newblock \emph{Nature} \textbf{2020}, \emph{579}, 7798 205.

\end{thebibliography}

\end{document}